\documentclass[]{sn-jnl}

\jyear{2023}%

\theoremstyle{thmstyleone}%
\theoremstyle{thmstyletwo}%

\theoremstyle{thmstylethree}%

\usepackage[version=3]{mhchem}
\usepackage{amsmath}
\usepackage{amssymb}

\usepackage{rotating}
\usepackage{subcaption}

\usepackage{numprint}
\usepackage{color}

\begin{document}

\title[ ]{Theory and simulation of shock waves freely propagating through monoatomic non-Boltzmann gas}


\author*[1]{\fnm{Malte} \sur{D\"{o}ntgen}}\email{doentgen@hgd.rwth-aachen.de}

\affil*[1]{\orgdiv{Chair of High Pressure Gas Dynamics, Shock Wave Laboratory}, \orgname{RWTH Aachen University}, \orgaddress{\city{Aachen}, \postcode{52056}, \country{Germany}}}


\abstract{
The effect of non-Boltzmann energy distributions on the free propagation of shock waves through a monoatomic gas is investigated via theory and simulation.
First, the non-Boltzmann heat capacity ratio $\gamma$, as a key property for describing shock waves, is derived from first principles via microcanonical integration.
Second, atomistic molecular dynamics simulations resembling a shock tube setup are used to test the theory.
The presented theory provides heat capacity ratios ranging from the well-known $\gamma = 5/3$ for Boltzmann energy-distributed gas to $\gamma \to 1$ for delta energy-distributed gas.
The molecular dynamics simulations of Boltzmann and non-Boltzmann driven gases suggest that the shock wave propagates about 9\;\% slower through the non-Boltzmann driven gas, while the contact wave appears to be about 4\;\% faster if it trails non-Boltzmann driven gas.
The observed slowdown of the shock wave through applying a non-Boltzmann energy distribution was found to be consistent with the classical shock wave equations when applying the non-Boltzmann heat capacity ratio.
These fundamental findings provide novel insights into the behavior of non-Boltzmann gases and might help to improve the understanding of gas dynamical phenomena.
}

\keywords{Microcanonical, Heat Capacity Ratio, Atomistic Molecular Dynamics, Non-Boltzmann Shock Wave}



\maketitle

\section{Introduction}\label{sec1}

%%%

% introduction of the problem
Non-equilibrium gas dynamics are essential for describing atmospheric entry processes~\cite{Boyce1996, Kosareva2021}, hypersonic cruise~\cite{Doroshenko1992}, and nozzle flow~\cite{Louviot2015, Dudas2020}.
% general introduction and short outline of paper focus
The mathematical formulation of such processes is well-established through the Boltzmann equation~\cite{Cercignani1985, Huang1987}, which allows for a continuous description of, typically, rarefied gas dynamics.
Non-Boltzmann dynamics of dense gases are commonly modeled through approximate schemes, many of which use the assumption of microcanonical equilibrium.
The present work provides a microcanonical non-equilibrium derivation of the heat capacity ratio of non-Boltzmann energy-distributed monoatomic gas.
Atomistic molecular dynamics simulations of shock waves freely propagating through Boltzmann and non-Boltzmann energy-distributed monoatomic gases are utilized to validate the theoretical findings.
Ultimately, the simulation results are compared to ideal equilibrium shock calculations using non-equilibrium heat capacity ratios.

% Boltzmann equation
The non-equilibrium state of a monoatomic gas is characterized by variations of the number density over the entire accessible phase space, spanned by the positions $q$ and the momenta $p$.
The Boltzmann equation provides the means for continuously describing the evolution of such a non-equilibrium state by phase space integration~\cite{Huang1987}.
Solving the Boltzmann equation, however, becomes increasingly demanding with decreasing Knudsen number, hence it is typically applied only for rarefied gas flows~\cite{Cercignani1985}.
Probably the most popular approach for averting the high computational efforts required for solving the Boltzmann equation is the use of Direct Simulation Monte Carlo (DSMC), which was pioneered by Bird~\cite{Bird1994}.
For a good overview of relevant studies about and with the DSMC method, please refer to Gallies~\cite{Gallies2019}.

% transition regime
For the transition regime from large to small Knudsen numbers, usually more efficient, yet more approximate approaches are utilized~\cite{Torrilhon2016}.
For intermediate Knudsen numbers the use of moment equations has been established, for which Torrilon~\cite{Torrilhon2016} provided a good overview of how moments are defined and how these equations can be used for modeling non-equilibrium gas flows.
For the strong non-equilibrium start-up phase of a shock tube process for instance, Au et al.~\cite{Au2001} found that the use of moment equations allows for reliably describing the first moments of start-up, which cannot be captured by the Navier-Stokes equation.
Another example is provided in the recent special issue on "nonequilibrium thermodynamics" edited by V\'{a}n~\cite{Van2020}, in which \"{O}ttinger et al.~\cite{Oettinger2020} formulated moment equations for rarefied gases based on the Boltzmann equation, showing that the rotational extended thermodynamics (RET) approach~\cite{Mueller1998} is a special case of the general equation for the non-equilibrium reversible-irreversible coupling (GENERIC)~\cite{Grmela1997}, with both approaches being used for describing non-equilibrium gas dynamics.

% Boltzmann energy distribution
With increasing gas density, i.e. decreasing Knudsen number, the models used for describing non-equilibrum start using the assumption of Boltzmann energy-distributed gases.
Various approaches for describing non-equilibrium thermodynamics have been developed~\cite{Lebon2008}, such as superstatistics~\cite{Beck2003}, stochastic thermodynamics~\cite{Seifert2008}, and local equilibrium thermodynamics~\cite{Hafskjold1995}.
The above methodologies all aim at formulating an unified description of non-equilibrium thermodynamics~\cite{Van2020}, yet they do not necessarily include non-Boltzmann energy distributions, i.e. local non-equilibrium in the momentum space.
The most widely used approach for modeling microcanonical non-equilibrium gas states is the use of distinct temperatures for distinguishable degrees of freedom~\cite{Schwartz1952}.
For instance, Kosareva et al.~\cite{Kosareva2021} recently applied a four-temperatures kinetic model for modeling the vibrational relaxation of \ce{CO2} during atmospheric entry.
This approach can generally be used to separate excitation of translational, rotational, vibrational, and electronic degrees of freedom, yet it is still based on the concept of the Boltzmann energy distribution.

% non-Boltzmann energy distribution
Besides the research on the Boltzmann equation and its derivatives for describing non-Boltzmann energy distributions, the role of non-Boltzmann energy distributions has been intensely investigated in the field of theoretical chemical kinetics recently~\cite{Burke2015b, Goldsmith2015a}.
In 2015, Klippenstein and co-workers~\cite{Burke2015b, Goldsmith2015a} revived the topic of non-Boltzmann chemical kinetics, which had probably been initially observed in the 1970s and 1980s~\cite{Dryer1971, Dean1985, Westmoreland1986}.
Subsequent studies by Labbe et al.~\cite{Labbe2016, Labbe2017} and by D\"{o}ntgen et al.~\cite{Doentgen2017, Doentgen2017b} advanced the fundamental understanding of non-Boltzmann chemical kinetics.
In particular, the prompt dissociation reactions of radical species was found to significantly affect macroscopic properties, such as the flame speed~\cite{Labbe2017} and the ignition delay time~\cite{Wildenberg2022}.
The combination of non-Boltzmann gas dynamics and non-Boltzmann chemical kinetics was intensely investigated by Kustova and co-workers~\cite{Kunova2015, Kosareva2021, Kravchenko2022, Melnik2022}, who investigated not only the role of vibrational exciation on \ce{CO2} chemistry during atmospheric entry~\cite{Kosareva2021}, but also the behavior of \ce{N2}~\cite{Kunova2015}, \ce{O2}~\cite{Kravchenko2022}, and \ce{CO}~\cite{Melnik2022} in vibrationally excited gas dynamical flows.
In this context, the use of state-to-state dynamics is widely established for describing the transition between energy levels and chemical states~\cite{Kunova2015, Sakamura2003, Kadochnikov2020, Kosareva2021, Kravchenko2022, Melnik2022}.

% relevance of non-Boltzmann effects
Non-equilibrium effects are known to or might be affecting various gas dynamical experiments.
For instance, Campbell et al.~\cite{Campbell2017} compared calculated and experimentally derived thermodynamic variables for postshock states of vibrationally excited, i.e. non-Boltzmann, gases.
The authors highlighted that vibrational excitation can lead to deviations in temperatures of up to 8\;\%~\cite{Campbell2017}.
As a less definite example, Bedin~\cite{Bedin1998} found an anomalous alteration of the velocity of a shock wave traversing a non-equilibrium plasma, which was intensely re-investigated experimentally~\cite{Ganguly1997, Bityurin2000} and numerically~\cite{Poggie1999, Molevich2004}.
The controversial discussion of the underlying mechanism is not yet fully resolved~\cite{Zhou2016b}, indicating the high complexity of the underlying mechanism and non-equilibrium gas dynamics in general.

% present work
The present work will show that non-Boltzmann energy distributions affect the heat capacity ratio, as the most relevant gas dynamical property, even at high density conditions.
First, the heat capacity ratio of a non-Boltzmann energy-distributed monoatomic gas will be derived from first principles by applying and integrating a microcanonical formulation of the heat capacity ratio, weighted by prototypical energy distributions.
Second, the theoretical dependence of the heat capacity ratio on the energy distribution will be validated through atomistic molecular dynamics simulations of a monoatomic gas for the two presently limiting energy distributions: The Boltzmann and the delta energy distributions.
Third, the shock velocities obtained through the molecular dynamics simulations will be compared to the ideal equilibrium shock calculations, carried out with the theoretical heat capacity ratios of the Boltzmann and delta energy distributions.
The present work will provide fundamental insights into the role of non-Boltzmann energy distributions on the most fundamental gas dynamical property: The heat capacity ratio.

%%%%%%%%%%%%%%%%%%%%%%%%%%%%%%%%%%%%%%%%%%%%%%%%%%%%%%%
\section{Theory}

The microcanonical heat capacity ratio, i.e. isentropic coefficient, $\gamma$ of a monoatomic ideal gas with an arbitrary energy distribution is derived here.
In the present work, the Boltzmann energy distribution (B) will be used as reference of systems in thermal equilibrium and a delta function ($\delta$) will be used to represent the upper limiting case of non-Boltzmann energy distributions.
In essence, a delta energy distribution represents a system in which all particles have exactly the same energy.
Such an energy distribution has limited physical relevance, since any collisional interaction would disrupt the delta energy distribution.
However, it allows to quantify the upper limiting effect of non-Boltzmann energy distributions on the heat capacity ratio.

The general definition of the heat capacity ratio is given as follows~\cite{Atkins2010p64}:
\begin{equation}\label{eq:gamma_heatcapacities}
\gamma = \frac{C_{P,j}}{C_{V,j}} ,
\end{equation}
with $j$ identifying the energy distribution, $C_{P,j}$ being the the isobaric heat capacity of distribution $j$, and $C_{V,j}$ being the isochoric heat capacity of distribution $j$.
The heat capacity is typically defined as the second derivative of the Helmholtz free energy with respect to temperature.
Here, however, it is more suitable to describe the heat capacity through the energy fluctuations of the system.
In a Boltzmann energy-distributed system, the heat capacity is given as:
\begin{equation}\label{eq:heatcapacity}
C_{i,j} = k_\text{B} \cdot \frac{\langle E^2 \rangle_{i,j} - \langle E \rangle_{i,j}^2}{(k_\text{B} T)^2} ,
\end{equation}
with $i$ being $P$ or $V$, $k_\text{B}$ being Boltzmann's constant, $E$ being the energy, and $\langle ... \rangle$ being the microcanonical averaging operator.
When combining equations \ref{eq:gamma_heatcapacities} and \ref{eq:heatcapacity}, one obtains the following expression:
\begin{equation}\label{eq:gamma_energies}
\gamma = \frac{\langle E^2 \rangle_{P,j} - \langle E \rangle_{P,j}^2}{\langle E^2 \rangle_{V,j} ,
 - \langle E \rangle_{V,j}^2}
\end{equation}
Since equation~\ref{eq:gamma_energies} solely depends on the averaged squared and squared averaged energies, $\langle E^2 \rangle$ and $\langle E \rangle^2$, respectively, it is applicable for arbitrary energy distributions.
The averaged squared energy $\langle E^2 \rangle$ is given as:
\begin{equation}\label{eq:ave_E2}
\langle E^2 \rangle_{i,j} = \frac{\int\limits_0^\infty E^2 \cdot \rho_i(E) \cdot f_j(E) \cdot \text{d}E}{\int\limits_0^\infty \rho_i(E) \cdot f_j(E) \cdot \text{d}E} ,
\end{equation}
with $\rho_i(E)$ being either the isobaric density of states ($i = P$) or the isochoric density of states ($i = V$) and $f_j$ being the propabilities of energy distribution $j$.
The squared averged energy $\langle E \rangle^2$ is given as:
\begin{equation}\label{eq:ave2_E}
\langle E \rangle_{i,j}^2 = \left( \frac{\int\limits_0^\infty E \cdot \rho_i(E) \cdot f_j(E) \cdot \text{d}E}{\int\limits_0^\infty \rho_i(E) \cdot f_j(E) \cdot \text{d}E} \right)^2 .
\end{equation}

For a monoatomic gas, the density of states is obtained through inverse Laplace transformation of the translational partition function~\cite{Doentgen2016b} and is independent of the energy distibution.
In case an isobaric state is considered, the density of states is given as:
\begin{equation}\label{eq:rho_P}
\rho_P(E) = \frac{q_\text{tr} / p}{\Gamma(5/2)} \cdot (E)^{3/2} ,
\end{equation}
with the translational partition function pre-factor $q_\text{tr} = (2 \pi m / h^2)^{3/2}$, the pressure $p$, and the gamma function $\Gamma$.
In case an isochoric state is considered, the density of states is given gas:
\begin{equation}\label{eq:rho_V}
\rho_V(E) = \frac{q_\text{tr} \cdot V}{\Gamma(3/2)} \cdot (E)^{1/2} ,
\end{equation}
with the volume $V$.

Combining equations~\ref{eq:gamma_energies} to \ref{eq:rho_V} yields the heat capacity ratio $\gamma$ for a given energy distribution $f_j(E)$.
For the Boltzmann energy distribution $f_B(E) = \exp(-E / k_\text{B} T)$ and the delta energy distribution $f_\delta(E) = \delta(E - E_0)$, $\gamma$ can be obtained analytically and amounts to $\gamma = 5/3$ and $\gamma \to 1$, respectively.
To allow for comparability between different energy distributions, the total energy of the compared energy distributions is required to be equal.
For the delta and Boltzmann energy distributions, the following equation relates the delta peak position $E_0$ to the temperature $T$ of the Boltzmann energy distribution.
\begin{equation}
 %\resizebox{0.5\textwidth}{!}{$
 \begin{split}
  \langle E_\delta \rangle &\overset{!}{=} \langle E_\text{B} \rangle \\
  \Leftrightarrow \frac{\int\limits_0^\infty E \cdot \rho (E) \cdot f_\delta(E) \cdot \text{d}E}{\int\limits_0^\infty \rho (E) \cdot f_\delta(E) \cdot \text{d}E} &= \frac{\int\limits_0^\infty E \cdot \rho (E) \cdot f_\text{B}(E) \cdot \text{d}E}{\int\limits_0^\infty \rho (E) \cdot f_\text{B}(E) \cdot \text{d}E} \\
  \Leftrightarrow \frac{\int\limits_0^\infty E \cdot \sqrt{E} \cdot \delta(E - E_0) \cdot \text{d}E}{\int\limits_0^\infty \sqrt{E} \cdot \delta(E - E_0) \cdot \text{d}E} &= \frac{\int\limits_0^\infty E \cdot \sqrt{E} \cdot \exp\left(-\frac{E}{k_\text{B} T}\right) \cdot \text{d}E}{\int\limits_0^\infty \sqrt{E} \cdot \exp\left(-\frac{E}{k_\text{B} T}\right) \cdot \text{d}E} \\
  \Leftrightarrow E_0 &= 3/2 \cdot k_\text{B} T
 \end{split}
 %$}
\end{equation}
The transition between the Boltzmann and delta energy distributions is modeled through the Gaussian energy distribution $f_\text{G}(E) = \frac{1}{\sqrt{2 \pi} \cdot b} \cdot \exp\left(-\frac{(E - E_{\text{G},0})^2}{2 b^2}\right)$ with the Gaussian width $b$ and the Gaussian position $E_{\text{G},0}$.
When setting $\langle E_\text{G} \rangle \overset{!}{=} \langle E_\text{B} \rangle$, the Gaussian width $b$ can be determined numerically for a given Gaussian position $E_{\text{G},0}$.
Through this procedure, the Gaussian distribution can be converged to the Boltzmann distribution for $E_{\text{G},0} \to \text{-}\infty$ and to the delta distribution for $E_{\text{G},0} \to 3/2 \cdot k_\text{B} T$.
For the Gaussian energy distributions, however, $\gamma$ cannot be determined analytically and the integrals in equations~\ref{eq:ave_E2} and \ref{eq:ave2_E} have to be solved numerically.

%%%%%%%%%%%%%%%%%%%%%%%%%%%%%%%%%%%%%%%%%%%%%%%%%%%%%%%
\section{Methodology}
% Lennard-Jones simulations of shock wave
Pseudo three-dimensional Lennard-Jones (LJ) molecular dynamics (MD) simulations have been carried out with super-critical Argon in the driver and driven sections at an initial temperature of 300\;K and ideal gas pressures of 500 and 50\;bar, respectively, using the LAMMPS software package~\cite{Plimpton1995}.
The Lennard-Jones potential used in the present MD simulations is defined as follows~\cite{LennardJones1931}:
\begin{equation}
V_\text{LJ} (r) = 4 \cdot \varepsilon \cdot \left ( \left(\frac{\sigma}{r}\right)^{12} - \left(\frac{\sigma}{r}\right)^6 \right) ,
\end{equation}
with distance $r$ between two LJ particles, LJ well depth $\varepsilon$, and LJ diameter $\sigma$.
The large pressures in the driver and driven sections are selected to generate a sufficiently strong shock wave for shock position post-processing and to keep the width of the shock wave rather small.
Thermodynamic real gas effects will be considered in the theoretical shock velocity calculations, as detailed below.
Argon is modeled via the LJ well depth $\varepsilon = 114\;\text{K}$ and the LJ diameter $\sigma_{3\text{D}} = 3.47\;\text{\AA}$.
To reduce the simulation size, the three-dimensional domain has been projected onto a two-dimensional domain and the LJ diameter has been adopted by requiring that the ideal gas LJ collision frequencies~\cite{Carstensen2007} of the three- and two-dimensional domains are equal, yielding $\sigma_{2\text{D}} = 2 \cdot \sigma_{3\text{D}}^2 / L$, with $L$ being the reference length of the reduced dimension.
Here, $L = 5\;\text{\AA}$ has been used.

The two-dimensional simulation box was $1\;\text{\textmu}\text{m}$ long and $0.2\;\text{\textmu}\text{m}$ wide.
The area defined by the first $0.1\;\text{\textmu}\text{m}$ of the long axis was filled with 11941 driver gas molecules, the remaining area was filled with 10852 driven gas molecules.
All molecules have been randomly positioned inside the respective areas.
Both ends of the short axis were modeled with periodic boundary conditions, through which the corresponding dimension of the simulation was pseudo-infinite.
The boundaries at both ends of the long axis were modeled as perfectly reflecting walls, through which collisions of gas particles with these walls were perfectly elastic.
The driver and driven sections were initially separated by five layers of particles with frozen positions.
Before removing the separating particle layers, the particle velocities were randomly distributed so that they resemble a Maxwell-Boltzmann distribution and each section was thermalized at 300\;K for 0.1\;ns using a Nos\'{e}-Hoover thermostat.
Then, the separating particle layers were removed and the entire system was simulated for 2\;ns without thermostating.
Post-processing, however, was stopped before the shock wave reached the endwall of the driven section.
A total of five repetitions with different random number generator seeds have been carried out for the Boltzmann and the delta energy-distributed driven gases.

The propagation of the contact wave between the driver and driven gases and the shock wave inside the driven gas was tracked through the particle density over the long axis of the simulation box.
The driven gas density was fitted using two logistic functions, one for modeling the contact wave and the other for modeling the incident shock wave.
The positions of the transitions from the lower to the upper limits of the logistic functions are used as contact and shock wave positions in the post-processing.
%The widths of the transition in the logistic functions are used to determine the width of the contact and shock waves.

Simulation of the Boltzmann energy-distributed driven gas was immediately possible, since the driven gas had been thermalized through the Nos\'{e}-Hoover thermostat and the subsequent constant particle (N), constant volume (V), and constant energy (E) simulations preserved the Boltzmann energy distribution in the driven gas.
A delta energy-distributed gas, however, would rapidly thermalize towards a Boltzmann energy-distributed gas at the present high-pressure conditions, which come with large collision frequencies.
Any collision within the driven gas would re-distribute the kinetic energy and relax the energy distribution towards the Boltzmann distribution.
This relaxation is especially pronounced for the delta energy distribution, as can be concluded from the H-function presented in Figure~\ref{fig:H_func}.
In order to limit the time available for relaxation of the delta energy distribution prior to arrival of the shock front at a specific point in the driven section, the delta energy distribution was actively maintained in the present simulations.
Based on a readily available thermostat routine, a novel routine has been implemented in the LAMMPS software package which rescales the velocities of the driven gas particles to a target velocity, which was defined so that the driven gas kinetic energies of the Boltzmann and delta energy distributions are equal.
To allow unbiased interactions of the driven gas with the shock front, the routine was applied to a dynamically shrinking sub-section of the driven section.
This shrinking sub-section spanned the entire short axis of the driven section, had one side fixed to the endwall of the driven section, and had the other side move towards the endwall at a constant velocity.
The shrinking velocity was the shock velocity obtained through the simulations of the Boltzmann energy-distributed gas, which was larger than the shock velocity obtained through the simulations of the non-Boltzmann energy-distributed gas.
Through this, a buffer zone between the sub-section in which the delta energy distribution was actively maintained and the shock front was implemented.

% checked with cantera + shock and detnonation toolbox
For comparison, the ideal shock velocities for the above simulated conditions were calculated using the Shock- and Detonation Toolbox~\cite{Shepherd2014} in conjunction with the Cantera software package~\cite{Goodwin2016}.
The effective pressures of the driver and driven sections used for shock velocity calculation were obtained through the Van-der-Waals equation of state with a molar co-volume of 0.03201\;l/mol and a co-pressure of $135.5\;\text{Pa}\cdot\text{l}^2 / \text{mol}^2$~\cite{EngineeringToolbox_VdW}.
This leads to effective pressures of 821 and 48\;bar for the driver and driven sections, respectively.
The pronounced real gas effects leading to the large driver pressure are acceptable in the present simulations, since the real gas interactions in the driver gas are not affecting the shock velocity in the driven section.
This shock velocity only depends on the ratio of effective pressures between driver and driven sections, the initial temperature, the molar mass, and the heat capacity ratio of the driven gas.
Since the driven gas is sufficiently well described by the ideal gas law (4\;\% deviation from ideal gas pressure), the presently calculated heat capacity ratios of the monoatomic ideal gases with a Boltzmann and a delta energy distribution can be used to determine the shock velocity in the driven section.
These shock velocities will be used for validating the present molecular dynamics simulations of Boltzmann energy-distributed gas and for testing if the same theories are applicable to non-Boltzmann energy-distributed gases.

%%%%%%%%%%%%%%%%%%%%%%%%%%%%%%%%%%%%%%%%%%%%%%%%%%%%%%%
\section{Results and Discussion}
\subsection{Non-Boltzmann Heat Capacity Ratio}
The theoretical heat capacity ratio of non-Boltzmann energy-distributed monoatomic gases provided above is evaluated numerically for a Gaussian energy distribution with the limiting cases of the Boltzmann and delta energy distributions.
As mentioned above, the Gaussian energy distribution converges to the Boltzmann energy distribution if the Gaussian position $E_{\text{G}, 0}$ diverges to $-\infty$ and the average energies of the two distributions are set to be equal.
For the sake of compact representation, however, the ratio of the isochoric averaged squared energies (cf. equations~\ref{eq:ave_E2} and \ref{eq:rho_V}) of the Boltzmann and Gaussian distributions $\langle E_\text{B}^2 \rangle / \langle E_\text{G}^2 \rangle$ is used to characterize the deviation from the Boltzmann distribution.
This ratio yields unity in case the Gaussian distribution resembles the Boltzmann distribution and it yields $15/9 \approx 1.67$ in case the Gaussian distribution resembles the delta distribution.

\begin{figure}[!htb]
 \centering
 \includegraphics[width=3in,keepaspectratio]{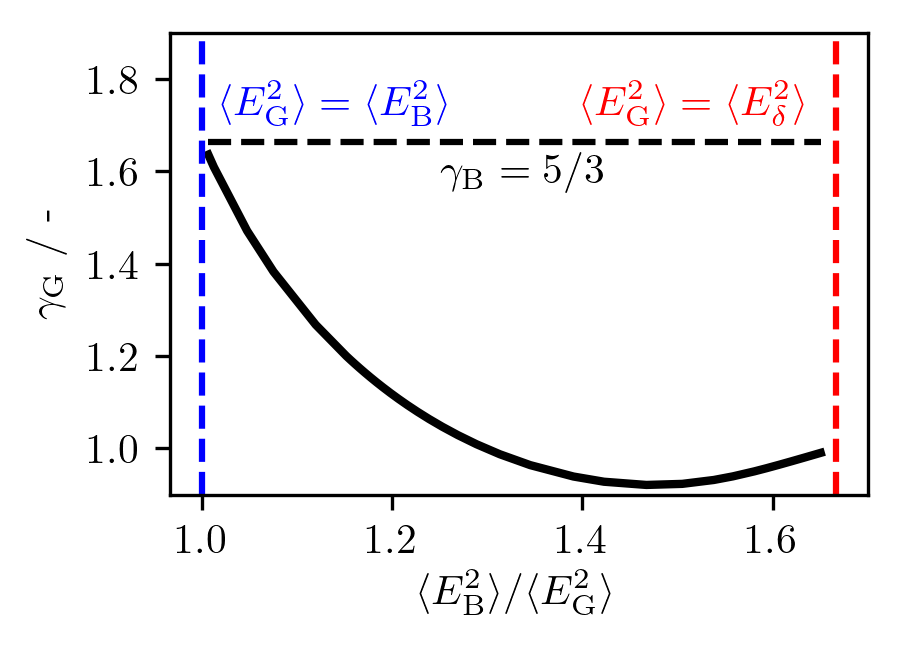}
 \caption{Heat capacity ratio $\gamma_\text{G}$ of a Gaussian energy-distributed monoatomic gas as function of the ratio of the averaged squared energies $\langle E^2 \rangle$ of the Boltzmann and Gaussian energy distributions.
 The horizontal dashed line represents the well-established heat capacity ratio of a Boltzmann energy-distributed monoatomic gas $\gamma_\text{B} = 5/3$.
 The vertical dashed lines represent the limiting cases defined through the Boltzmann and delta energy distributions, $\langle E_\text{G}^2 \rangle = \langle E_\text{B}^2 \rangle$ and $\langle E_\text{G}^2 \rangle = \langle E_\delta^2 \rangle$, respectively.}
 \label{fig:gamma}
\end{figure}

Figure~\ref{fig:gamma} shows the heat capacity ratio of a Gaussian energy-distributed monoatomic gas $\gamma_\text{G}$ as function of the ratio of the averaged squared energies of the Boltzmann and Gaussian distributions.
The dashed lines represent the Boltzmann heat capacity ratio (cf. black dashed line, $\gamma_\text{B} = 5/3$), the Boltzmann limiting case of the Gaussian distribution (cf. blue dashed line, $\langle E_\text{G}^2 \rangle = \langle E_\text{B}^2 \rangle$), and the delta limiting case of the Gaussian distribution (cf. red dashed line, $\langle E_\text{G}^2 \rangle = \langle E_\delta^2 \rangle$).

As would be expected, the Gaussian heat capacity ratio resembles the Boltzmann and delta heat capacity ratios if the Gaussian distribution converges to either of these two limiting distributions.
However, there is an unexpected minimum for $\langle E_\text{B}^2 \rangle / \langle E_\text{G}^2 \rangle \approx 1.47$, for which the heat capacity ratio of the respective Gaussian distribution amounts to $\gamma_\text{G} \approx 0.92$.
This averaged squared energy ratio corresponds to a Gaussian position rather close to the upper limiting case of $E_0 = 3/2 \cdot k_\text{B} T$, with $E_{\text{G},0} = 0.9 \cdot E_0$.
It was tested if the observed minimum can be attributed to numerical uncertainties, yet changing the convergence threshold and the energy increment of the numerical integrations from $10^{-9}$ to $10^{-15}$ and from $1\;\text{cm}^{-1}$ to $0.1\;\text{cm}^{-1}$, respectively, had a negligible effect on $\gamma$ in the range of $10^{-5}$.
This means that the numerically calculated $\gamma_\text{G}$ curve is an accurate representation of the underlying theory and that the observed $\gamma_\text{G} < 1$ is theoretically valid.

From a classical thermodynamics perspective, the heat capacity ratio determines the volume response to adiabatic temperature changes as $\text{d}V/\text{d}T = (C_\text{p} - C_\text{V}) / p$.
In this classical context, a heated gas with $\gamma > 1$ would expand, with $\gamma = 1$ its volume would not change, and with $\gamma < 1$ it would contract.
In the present work, the effect of non-Boltzmann energy distributions on the adiabatic change of states was tested by randomly sampling velocities from Gaussian energy distributions and optimizing the Gaussian peak position $E_{\text{G}, 0}$ and width $b$ so that the pressure, the internal energy, and the entropy remained constant.
All presently tested Gaussian energy-distributed monoatomic gases expand with increasing temperature according to the well-established description of isentropic expansion within numerical uncertainties: $V_2 / V_1 = (T_1 / T_2)^{1 / (\gamma - 1)}$, with $\gamma = 5/3$.
This means that the non-Boltzmann heat capacity ratio must not be confused with the classical heat capacity ratio for thermal equilibrium.

The observed $\gamma < 1$ originates from the energy dependencies of the isochoric and isobaric densities of states.
While the isochoric density of states increases sub-proportional with energy, the isobaric density of states increases super-proportional with energy.
As a consequence, the relative population of low energy states is larger and the relative population of high energy states is smaller for isochoric conditions compared to isobaric conditions.
While this still leads to $C_\text{P} > C_\text{V}$ for probability distributions which monotonically decline with energy, such as the Boltzmann distribution or the Gaussian distribution with $\langle E_\text{B}^2 \rangle / \langle E_\text{G}^2 \rangle < 1.22$, it results in $C_\text{P} \leq C_\text{V}$ for a range of non-monotone Gaussian distributions with $\langle E_\text{B}^2 \rangle / \langle E_\text{G}^2 \rangle \gtrapprox 1.3$.
For these Gaussian energy distributions, the contributions of the densely populated low energy states to the isochoric heat capacity exceed the contributions of the sparsely populated high energy states to the isobaric heat capacity.

\begin{figure}[!htb]
 \centering
 \includegraphics[width=3in,keepaspectratio]{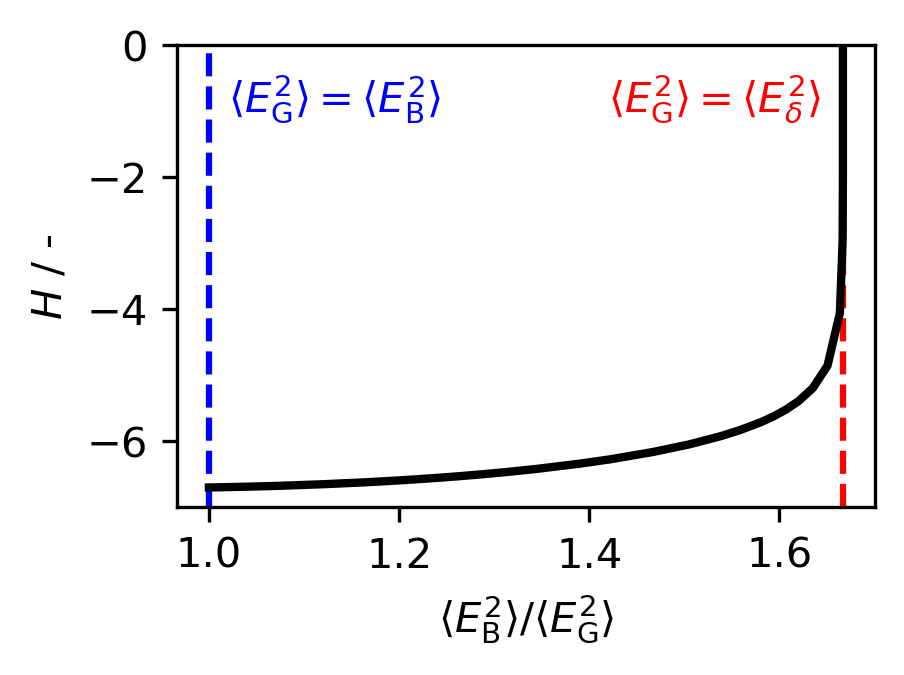}
 \caption{$H$-function of the Boltzmann, Gaussian, and delta energy distributions as function of the ratio of the averaged squared energies $\langle E^2 \rangle$ of the Boltzmann and Gaussian energy distributions.
 The vertical dashed lines represent the limiting cases defined through the Boltzmann and delta energy distributions, $\langle E_\text{G}^2 \rangle = \langle E_\text{B}^2 \rangle$ and $\langle E_\text{G}^2 \rangle = \langle E_\delta^2 \rangle$, respectively.}
 \label{fig:H_func}
\end{figure}

While the energy of the presently investigated probability distributions is set to be equal, the entropy is certainly not.
Here, Boltzmann's $H$-theorem~\cite{Boltzmann1872} is utilized to obtain insights into the stability / entropy of the Boltzmann, Gaussian, and delta energy distributions.
Figure~\ref{fig:H_func} shows the $H$-function of the Boltzmann energy distribution (left-most limit), of the delta energy distribution (right-most limit), and of the Gaussian energy distributions interpolating between these two limiting cases.
The H-function is calculated as follows:
\begin{equation}
 H = \sum\limits_k p_k \cdot \ln\left( p_k \right) ,
\end{equation}
with the probability $p_k$ of energy $E_k$, defined as $p_k = \rho_i(E_k) \cdot f_j(E_k) / \int\limits_0^\infty \rho_i(E) \cdot f_j(E) \cdot \text{d}E$.

As would be expected, the Boltzmann energy distribution comes with the smallest value of the $H$-function.
The delta energy distribution comes with a $H$-function value of zero, which indicates that the entropy of the delta energy distribution is also zero.
This underlines that the Boltzmann and delta energy distributions are suitable limits for the investigation of non-Boltzmann energy distributions.
Starting from the delta energy distribution, the H-function of Gaussian energy distributions rapidly drops from zero to a monotonous convergence towards the $H$-function value of the Boltzmann energy distribution.
The rapid drop of the $H$-function close to the delta limiting case indicates that the delta energy distribution is entropically extremely unstable.

\subsection{Atomistic Shock Wave Simulations}
% introduction mol dyn
The present atomistic molecular dynamics simulations are used to test and validate the above theoretical findings on the non-Boltzmann heat capacity ratio.
The simulations resemble an ideal shock tube process, as the heat capacity ratio directly affects the velocity at which shock waves propagate through the driven gas.
First, potential non-idealities of the simulations will be discussed, followed by discussion of the simulation results and their comparison to theory.

\begin{figure}[!htb]
 \centering
 \includegraphics[width=3.3in, keepaspectratio]{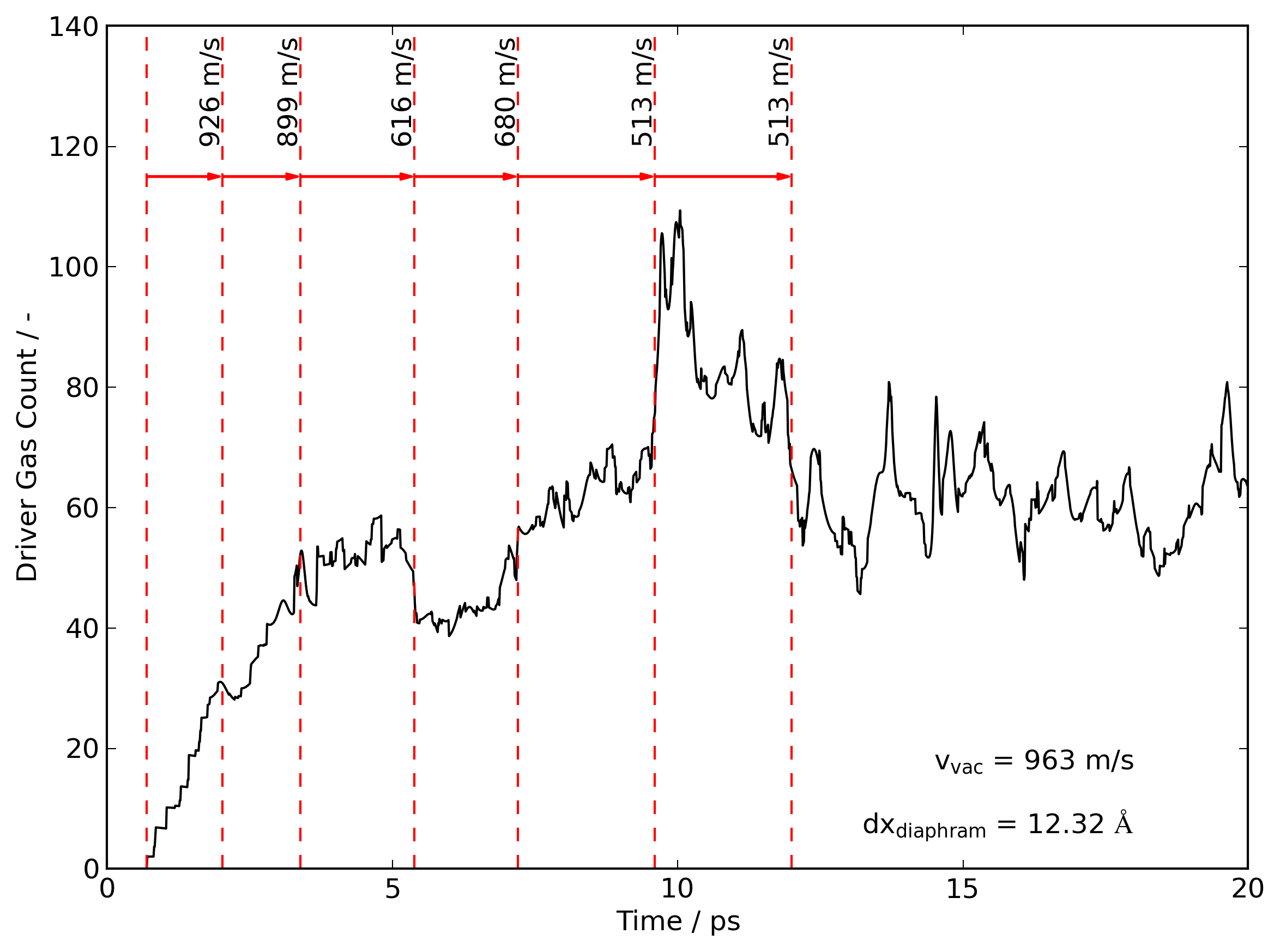}
 \caption{Driver gas count during the inception phase directly after abrupt removal of the diaphragm particles which separated the driver and driven sections during thermalization.
	The theoretical vacuum expansion velocity~\cite{Greenspan1962} for the simulated condition is provided as reference.}
 \label{fig:DiaphragmShocks}
\end{figure}

% initialization phase (diaphragm removal and related shocks
Fixed particles have been used as diaphragm between the driver and driven sections of the simulation to allow independent thermalization of the two gases.
The abrupt removal of the diaphragm particles, however, leaves a 12.32~\AA{} wide void between the two sections.
Based on the rather large effective pressure ratio between the driver and driven sections of about 17 it is assumed that the void between the driver and driven sections is solely penetrated by the driver gas.
This penetration through the driver gas into the void generates expansion and compression waves bouncing forth and back within this void during the first 12\;ps of the simulation, which have been traced through the driver gas count.
Figure~\ref{fig:DiaphragmShocks} shows an exemplary driver gas count during the inception phase of the main simulation directly after abruptly removing the diaphragm particles.

When loosely attributing kinks and jumps in the driver gas count to expansion and compression waves, one can find three characteristic velocities: 899-926\;m/s, 616-680\;m/s, and 513\;m/s.
Each characteristic velocity comes in pairs, indicating that each wave which traverses the void towards the driven section is being reflected at the interface between the void and the driven section.
With each expansion and compression process, the void is filled with more driver gas particles, leading to a reduction in the characteristic velocities of the succeeding waves.
While the very first forth and back propagating wave has a characteristic velocity comparable to the theoretical vacuum expansion velocity $v_\text{vac} = 963\;\text{m}/\text{s}$ based on the formula provided by Greenspan and Butler~\cite{Greenspan1962}, the characteristic velocities of the two succeeding forth and back propagating waves are already rather close to the shock velocity of the main shock wave propagating through the bulk of the driven gas after the inception phase ($v_\text{S} = 545.56\;\text{m}/\text{s}$ for the particular example in Figure~\ref{fig:DiaphragmShocks}, cf. Table~\ref{tab:ShockVelocities} for all resutls).

\begin{figure*}[!htb]
 \centering
 % Boltzmann
 \begin{subfigure}[b]{2.0in}
 \centering
 \subcaption{Boltzmann}
 \label{fig:Boltzmann_Sequence}
 \includegraphics[angle=0, height=0.21\textheight, keepaspectratio]{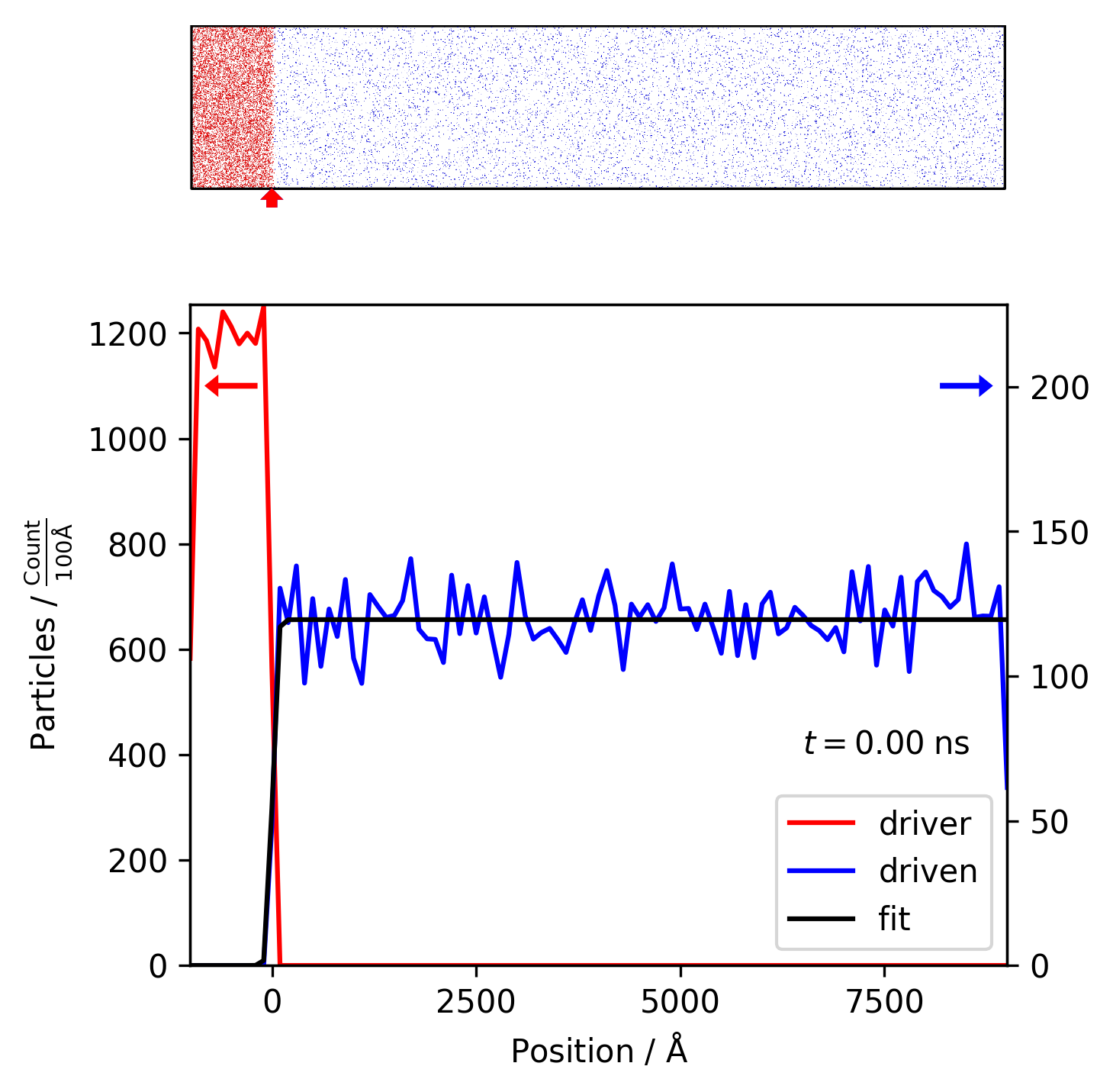}\\
 \includegraphics[angle=0, height=0.21\textheight, keepaspectratio]{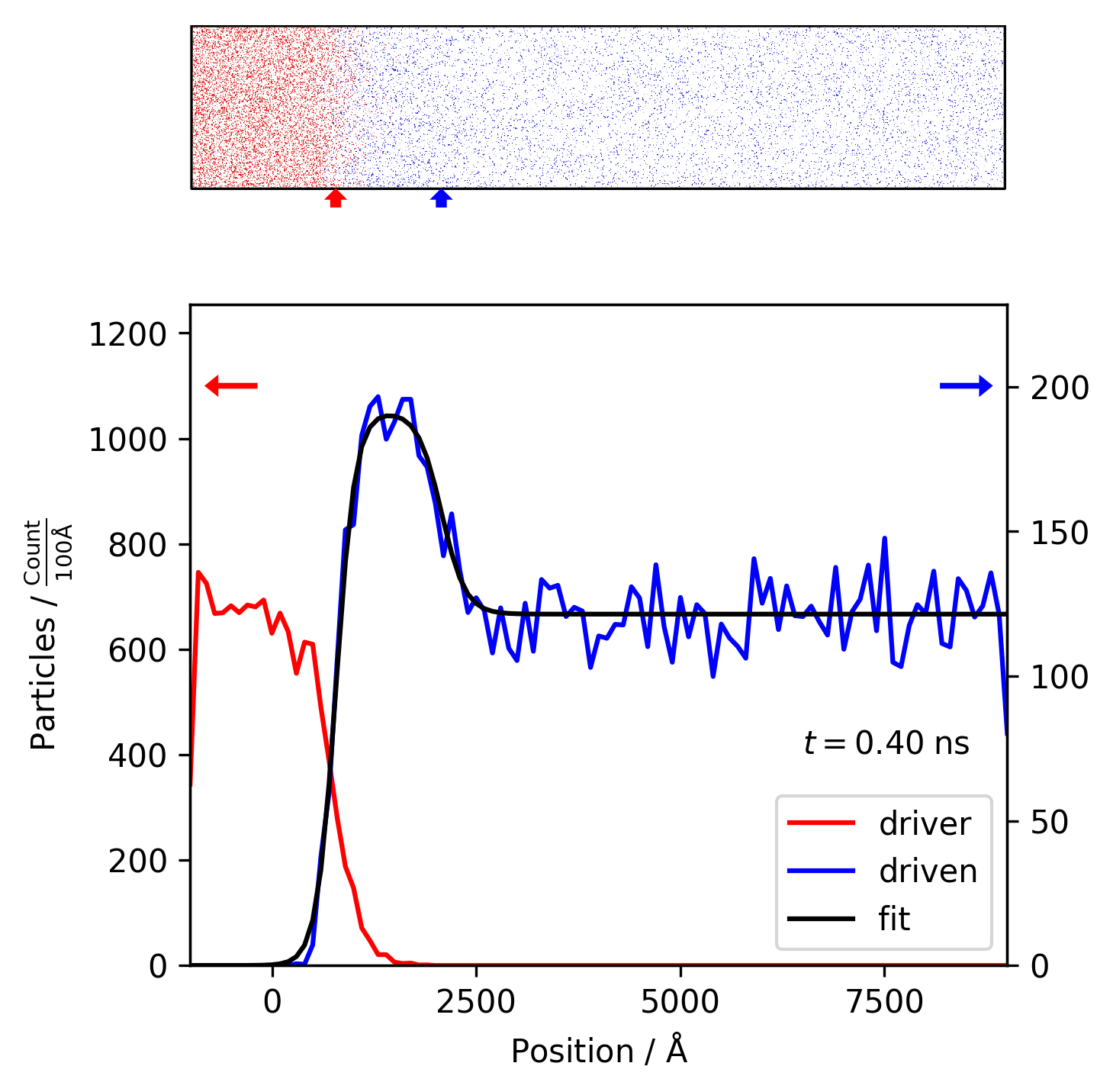}\\
 \includegraphics[angle=0, height=0.21\textheight, keepaspectratio]{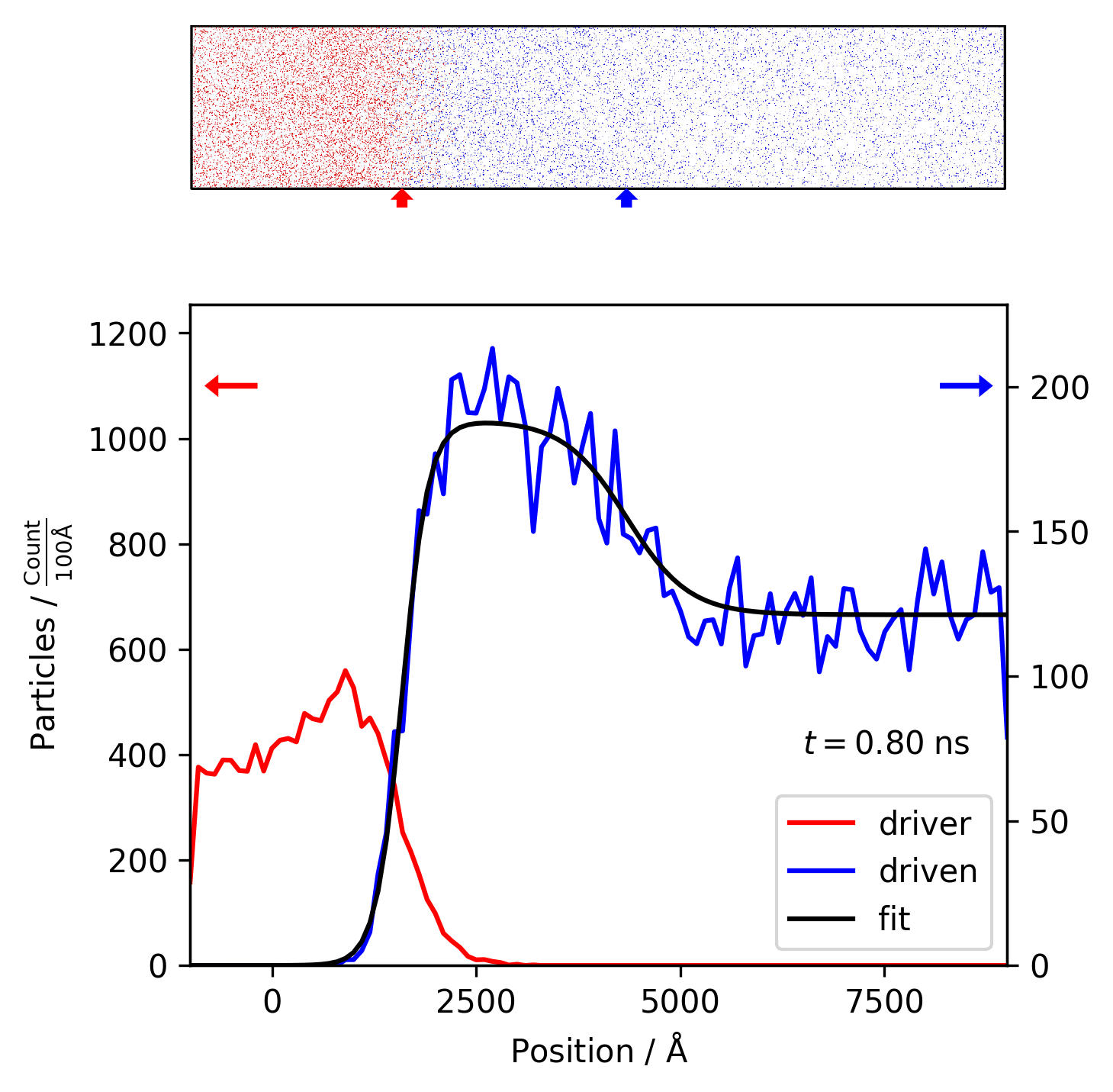}\\
 \includegraphics[angle=0, height=0.21\textheight, keepaspectratio]{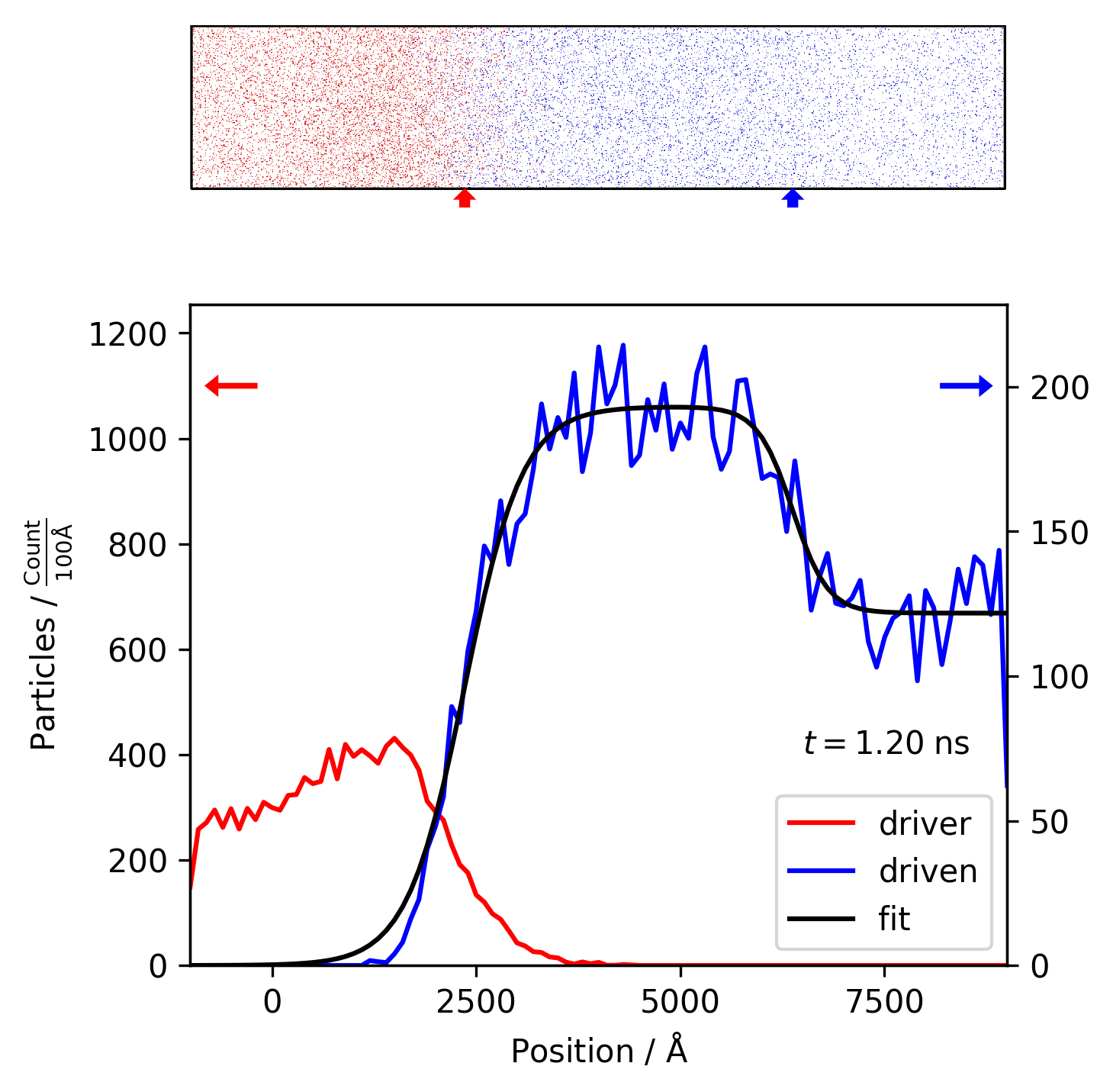}
 \end{subfigure}%
 % non-Boltzmann
 \begin{subfigure}[b]{2.0in}
 \centering
 \subcaption{non-Boltzmann}
 \label{fig:nonBoltzmann_Sequence}
 \includegraphics[angle=0, height=0.21\textheight, keepaspectratio]{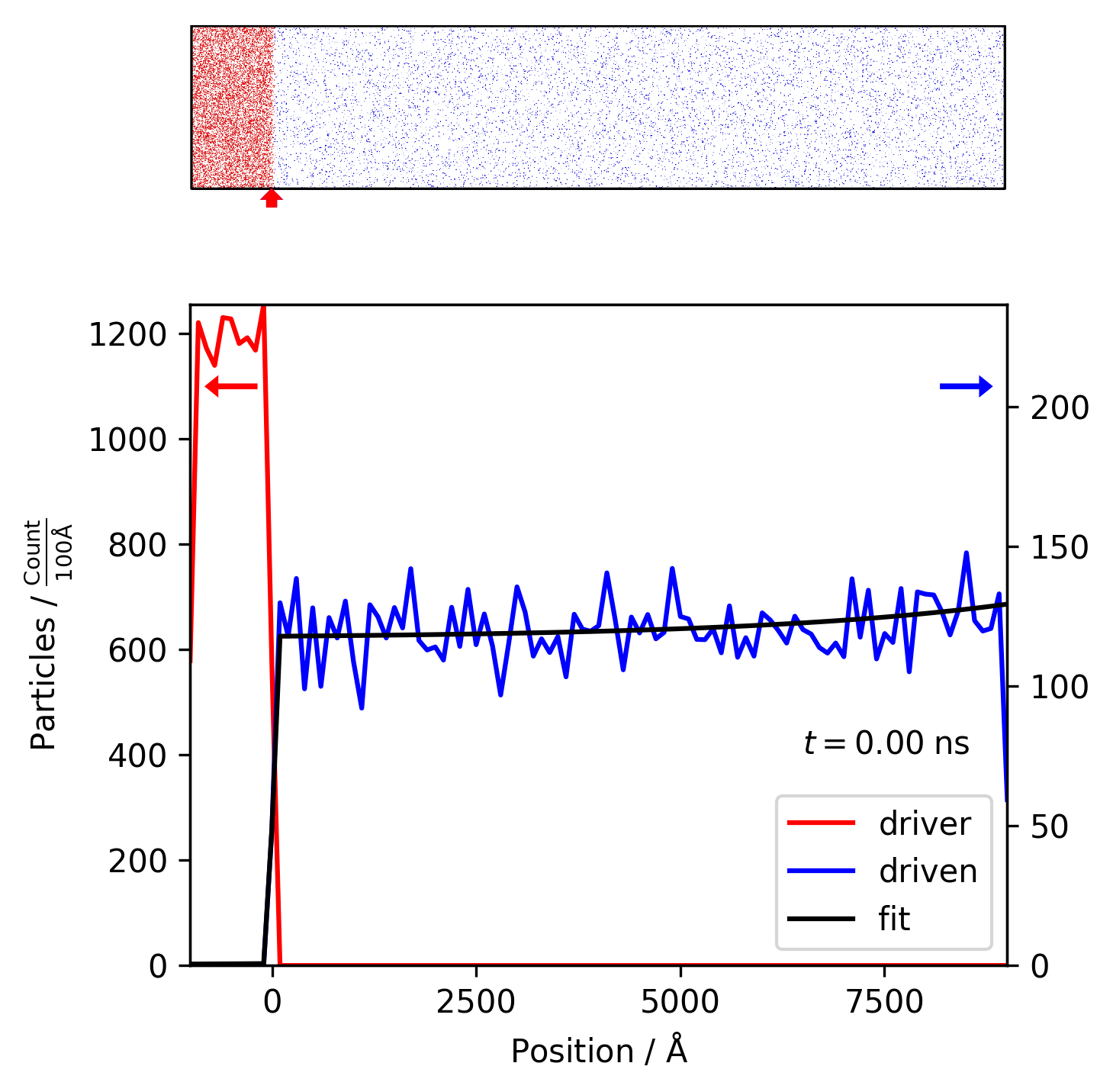}\\
 \includegraphics[angle=0, height=0.21\textheight, keepaspectratio]{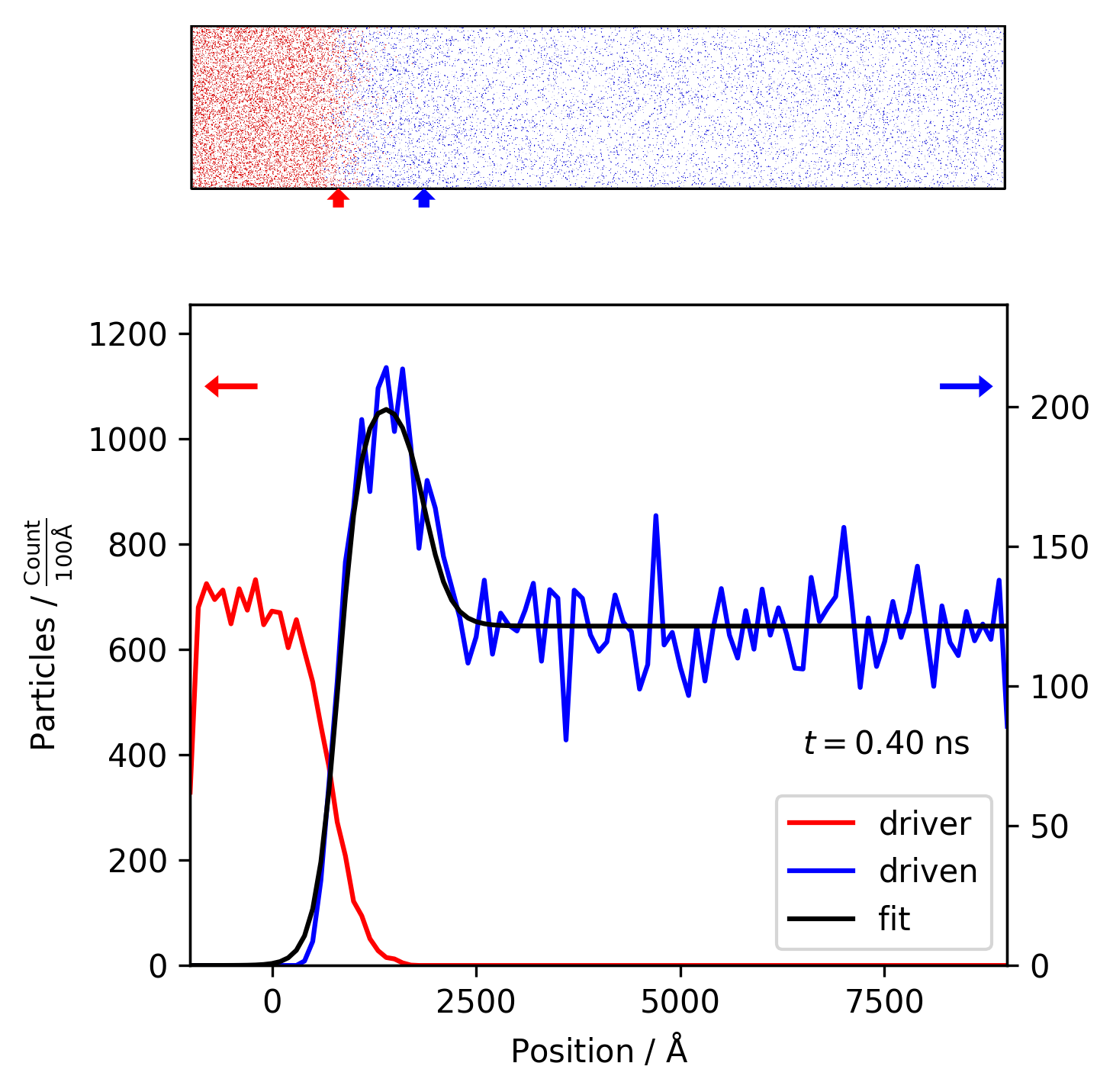}\\
 \includegraphics[angle=0, height=0.21\textheight, keepaspectratio]{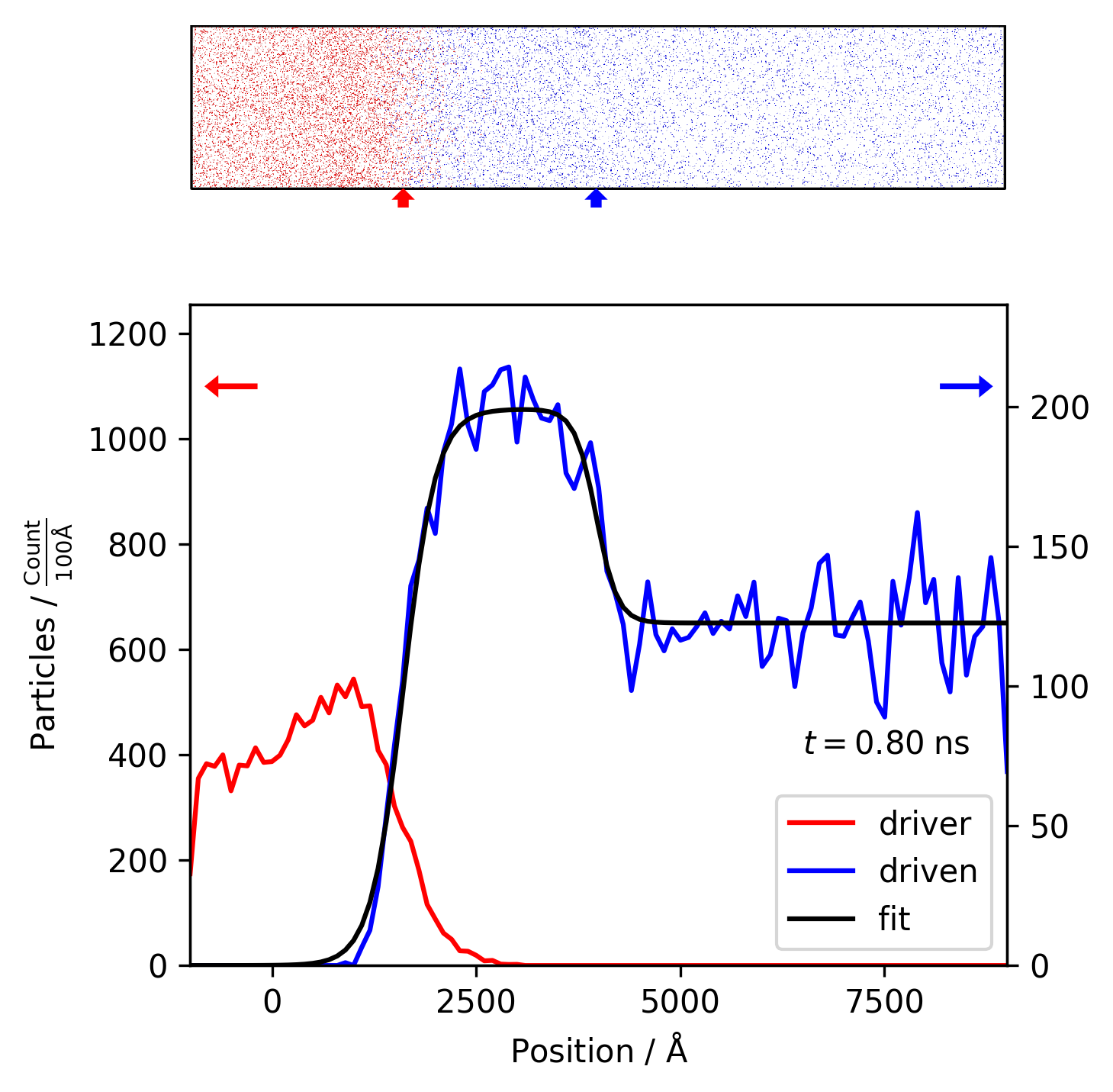}\\
 \includegraphics[angle=0, height=0.21\textheight, keepaspectratio]{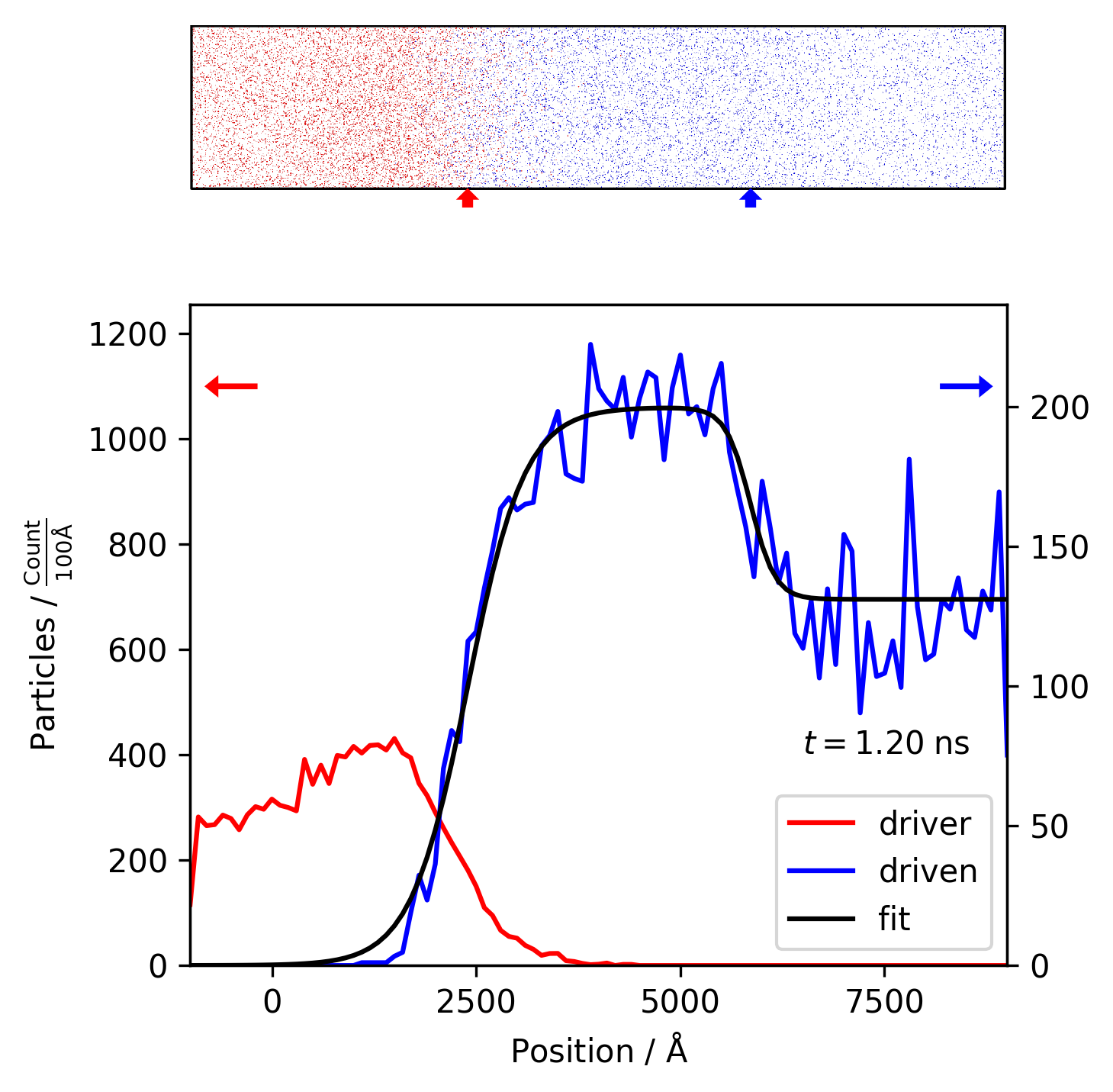}
 \end{subfigure}
 \caption{Sequences of frames taken from the post-processing of the simulations of Boltzmann and non-Boltzmann energy-distributed driven gases with the same random number generator seed.}
 \label{fig:Sequences}
\end{figure*}

After the third forth and back propagating wave has traversed the void between the driver and driven sections, the driver gas flow is established and the driver gas count remains constant at the time scale relevant for the inception phase.
The large  volume ratio of the driven section to the driver section of 9, however, causes a significant drop in driver gas concentration at later times, as visible in the sequence of frames provided in Figure~\ref{fig:Sequences}.
Neither the dropping pressure in the driver section nor the observed bouncing expansion and compression waves during the inception phase are expected to affect the shock velocity of the main shock wave freely propagating through the bulk of the driven gas.

% frame sequences
Each simulation is post-processed based on the driver and driven gas particle counts along the long axis of the simulation, which ranges from -1000\;\AA{} to 9000\;\AA.
In order to improve statistics, ten subsequent simulation steps are always merged into a single frame, yielding a total of \numprint{119410} driver gas particles and \numprint{108520} driven gas particles for evaluating the particle distribution along the x-axis.
Figure~\ref{fig:Sequences} shows two sequences of such particle distributions together with snapshots of the simulation boxes for a Boltzmann energy-distributed driven gas (\ref{fig:Boltzmann_Sequence}) and for a non-Boltzmann energy-distributed driven gas (\ref{fig:nonBoltzmann_Sequence}).
Note that both simulations have an identical random number generator seed, thus are identical except for the energy distribution of the driven gas preceding the shock wave.
The particle counts are discretized along the long axis in 100\;\AA{} increments with the driver gas particles shown in \color{red}{red}\color{black}{} and the driven gas particles shown in \color{blue}{blue}\color{black}{} for both the simulation snapshots and the particle count plots.
The red left pointing arrow and the blue right pointing arrow in the particle count plots assign the left and right y-axes of the plots to the driver and driven gas particle counts, respectively.
The upward pointing red and blue arrows directly below the simulation snapshots in turn indicate the contact and shock wave positions, respectively.
These positions were determined through the steps of the double sigmoid fits to the driven gas particle counts (black lines).
The full video of the presently provided simulation frames can be found in the supplementary information.

The first observation from Figure~\ref{fig:Sequences} is that the shock positions differ between the Boltzmann and non-Boltzmann energy-distributed driven gases.
In the Boltzmann case, the shock velocity extracted from the provided example is 545.56\;m/s, while the shock velocity in the corresponding non-Boltzmann case is only 477.18\;m/s.
Interestingly, the shock front appears to be thicker in the Boltzmann case compared to the non-Boltzmann case, with a sigmoid width of 70.12\;\AA{} for the Boltzmann case and 28.64\;\AA{} for the non-Boltzmann case, averaged over the entire simulations.
The differences in shock velocities and in shock front widths can be qualitatively explained by the differences in the energy distributions.
The presently applied non-Boltzmann energy distribution closely resembles a delta energy distribution, thus the driven gas particles have almost exclusively one randomly oriented velocity.
This means that the driven gas particles do not have high or low velocity particles, as it is the case for a Boltzmann energy-distributed driven gas.
The high velocity gas particles facilitate a faster shock propagation, leading to the observed larger shock velocity in the Boltzmann case.
At the same time, the low velocity particles are trailing the shock front, leading to a more stretched shock front in the Boltzmann case.

\begin{figure}[!htb]
 \centering
 \includegraphics[width=3.3in, keepaspectratio]{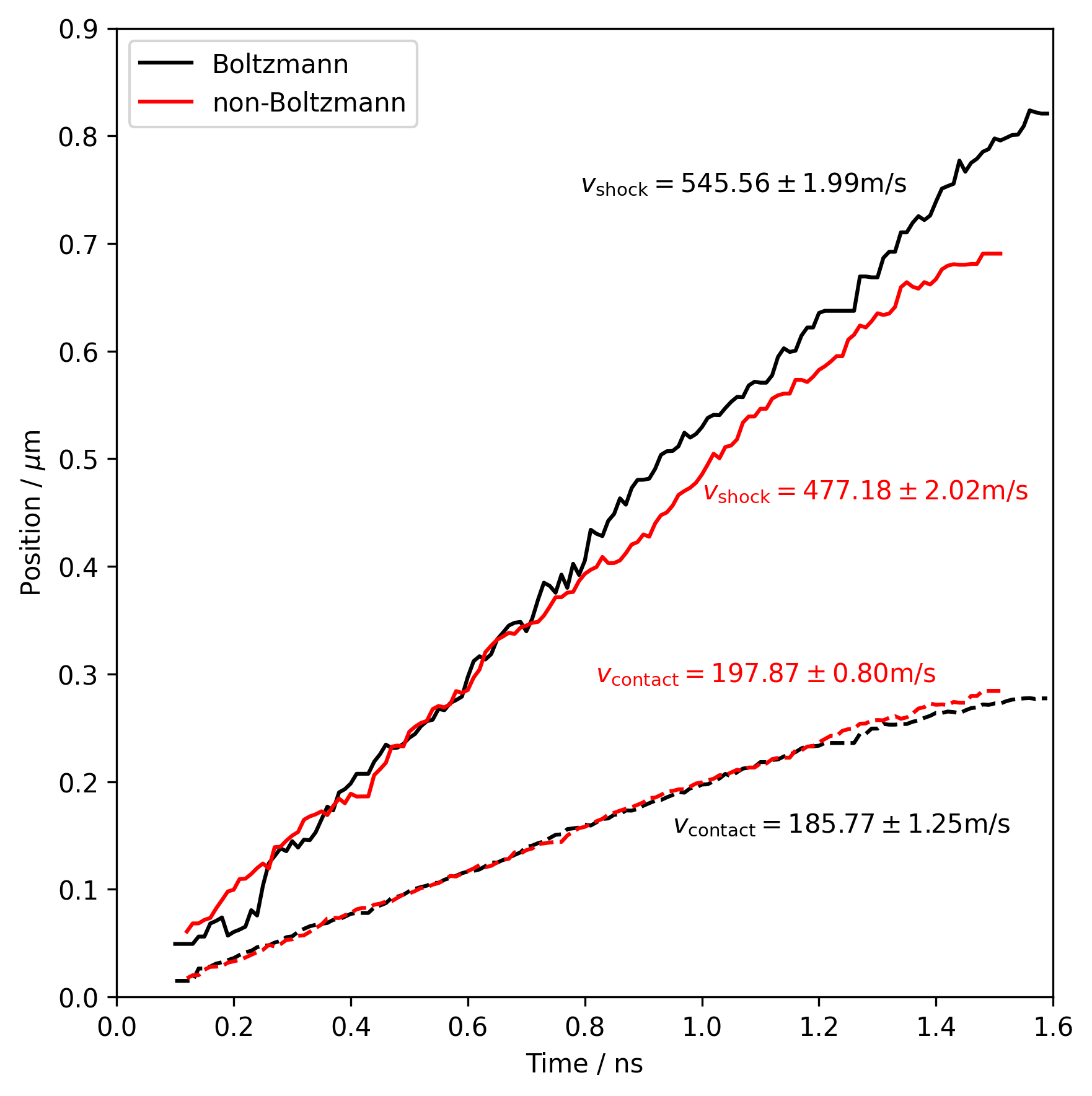}
 \caption{Contact and shock wave positions over simulation time for two exemplary simulations of a comparable Boltzmann and non-Boltzmann case.}
 \label{fig:ShockVelocityComparison}
\end{figure}

% wave propagation
For each replica simulation of the Boltzmann and the non-Boltzmann cases, the contact and shock wave positions are obtained as described above and traced over time.
Figure~\ref{fig:ShockVelocityComparison} shows the wave position profiles for the two exemplary simulations provided in Figure~\ref{fig:Sequences}.
The reported wave velocities and uncertainties are obtained via linear fitting, with the uncertainties being explicitly for the slope of the linear fit.

% tabulated results
This linear fitting procedure is carried out for all replica simulations and the resulting contact and shock wave velocities with their respective fitting uncertainties are tabulated in Table~\ref{tab:ShockVelocities}.
The wave velocities are sorted by the energy distribution case of the driven gas, either Boltzmann or non-Boltzmann.

\begin{table}
 \centering
 \caption{Shock wave velocities $v_\text{S}$ and contact wave velocities $v_\text{C}$ of the Boltzmann and non-Boltzmann driven gases.}
 \label{tab:ShockVelocities}
 \begin{tabular}{lllll}
  Replica & $v_{\text{S}, \text{sim}}$ & $v_{\text{S}, \text{sim}}$ err. & $v_{\text{C}, \text{sim}}$ & $v_{\text{C}, \text{sim}}$ err. \\
   & m/s & m/s & m/s & m/s \\
	\hline
	\multicolumn{5}{c}{\textbf{Boltzmann}} \\
	1 & 545.56 & 1.99 & 185.77 & 1.25 \\
	2 & 555.43 & 2.58 & 182.73 & 1.10 \\
	3 & 558.66 & 1.63 & 180.83 & 1.16 \\
	4 & 549.24 & 1.75 & 181.44 & 1.06 \\
	5 & 536.65 & 1.52 & 183.45 & 1.33 \\
	\multicolumn{5}{c}{\textbf{non-Boltzmann}} \\
	1 & 477.18 & 2.02 & 197.87 & 0.80 \\
	2 & 502.92 & 2.33 & 188.64 & 1.05 \\
	3 & 516.63 & 3.62 & 191.27 & 1.08 \\
	4 & 515.82 & 3.13 & 188.12 & 1.05 \\
	5 & 493.42 & 2.17 & 182.76 & 1.53 \\
	\hline
 \end{tabular}
\end{table}

The differences observed in Figure~\ref{fig:Sequences} for a single example of a Boltzmann and a non-Boltzmann case prevail in the other replica simulations, as well as in the averaged velocities.
The five simulations with the Boltzmann energy-distributed driven gas give an average shock velocity of $v_{\text{S}, \text{B}, \text{sim}} = 549.11\pm6.40\;\text{m}/\text{s}$ and the five simulations with the non-Boltzmann energy-distributed driven gas give an average shock velocity of $v_{\text{S}, \delta, \text{sim}} = 501.19\pm12.72\;\text{m}/\text{s}$.
This means that the presently employed non-Boltzmann energy distribution, which closely resembles the delta distribution, hinders the free propagation of shock waves through the monoatomic gas by about 9\;\% with respect to the shock velocity.
Note that convergence of the presented results has been tested by simulating with a doubled temporal resolution and post-processing with doubled spatial resolution, resulting in -0.08\;\% and -0.1\;\% deviations in the shock velocity, respectively.
In addition, the total energy of the present NVE simulations should be constant and was found to deviate by 0.13\;\% over the entire simulation time at most.

Therefore, the presented results are assumed to be sufficiently converged and the observed scatter is solely attributed to the statistical uncertainties inherent to molecular dynamics simulations~\cite{Kroeger2017}.
For the non-Boltzmann molecular dynamics simulation, the speed of the dynamically shrinking sub-section which is used to maintain the delta energy distribution has been modified as well, resulting in a shock velocity within the aforementioned methodological uncertainty.

Interestingly, the contact wave appears to accelerate from an average of $v_{\text{C}, \text{B}, \text{sim}} = 183.04\pm1.25\;\text{m}/\text{s}$ to $v_{\text{C}, \delta, \text{sim}} = 189.73\pm3.87\;\text{m}/\text{s}$ when applying a delta energy distribution to the driven gas.
This effect amounts to about 4\;\% and is larger than the observed statistical uncertainties of the simulations.
Noteworthy, this effect is trailing the incident shock wave, yet the gas behind the shock wave would typically be expected to be unaffected by the energy distribution in front of the shock wave.

% general agreement with theory
The averaged shock velocities of the Boltzmann and non-Boltzmann cases agree well with the ideal shock calculations using the incident shock prediction of the Shock- and Detonation Toolbox~\cite{Shepherd2014}, which yields $v_{\text{S}, \text{B}, \text{calc}} = 548\;\text{m}/\text{s}$ for $\gamma = 5/3$ and $v_{\text{S}, \gamma \to 1, \text{calc}} = 494\;\text{m}/\text{s}$ when slowly converging $\gamma$ to unity.
While the shock velocity is ill-defined for $\gamma = 1$, it converges with $\gamma \to 1$.
Although the value for $\gamma \to 1$ is slightly below the average simulation result, it is clearly within the statistical uncertainty of the simulations.
The comparison of the ideal shock calculation with the present simulations show that a non-Boltzmann heat capacity ratio can be combined with the standard description of ideal shock waves, which has originally been formulated for Boltzmann energy-distributed gases.
As mentioned above, however, the non-Boltzmann heat capacity ratio should not be confused with the classical heat capacity ratio for thermal equilibrium.

%%%%%%%%%%%%%%%%%%%%%%%%%%%%%%%%%%%%%%%%%%%%%%%%%%%%%%%
\section{Conclusions}
% theory
In the present study, a formulation of the heat capacity ratio of non-Boltzmann energy-distributed monoatomic gas has been derived from first principles.
This theory is applied to non-Boltzmann energy distributions which all exhibit the same total energy and are limited by the Boltzmann energy distribution as the equilibrium case and the delta energy distribution as the most extreme non-equilibrium case.
A continuous description of distributions between the two limiting cases is facilitated using a Gaussian distribution with variable peak position.
Applying the presented theory in the Boltzmann case yields the well-established heat capacity ratio of $\gamma_\text{B} = 5/3$, while in the delta case it yields $\gamma_\delta \to 1$.
Interestingly, applying the presented theory to the Gaussian case with a peak position close to the delta peak position, the heat capacity ratio appears to drop below unity.
Investigation of the isentropic thermal expansion behavior of non-Boltzmann energy-distributed monoatomic gas, however, revealed that the energy distribution does not affect thermal expansion and that a monoatomic gas always expands with a $\gamma = 5/3$ in the isentropic case.
To stress it once more, the non-Boltzmann heat capacity ratio should not be confused with the classical heat capacity ratio for thermal equilibrium.

% simulation
The presented atomistic molecular dynamics simulations resemble a shock tube process and have been carried out for Boltzmann and non-Boltzmann energy-distributed driven gases.
The contact wave and shock wave velocities extracted from these simulations reveal a consistent difference between the Boltzmann and non-Boltzmann cases.
While the shock wave propagates about 9\;\% slower through the non-Boltzmann driven gas, the contact wave appears to be faster by 4\;\% when trailing a shock wave through a non-Boltzmann driven gas.
The difference between the Boltzmann and non-Boltzmann shock wave velocities are reproduced through the ideal shock equations when utilizing the heat capacity ratios of $\gamma = 5/3$ and $\gamma \to 1$ for the Boltzmann and non-Boltzmann cases, respectively, as provided by the present theory.
Firstly, this validates the present theory for non-Boltzmann heat capacity ratios in the context of shock propagation and secondly, it appears that non-Boltzmann heat capacity ratios can be combined with the classical shock wave formulations for describing non-Boltzmann gas dynamics.

% outlook
Building on the presented theory, the non-Boltzmann heat capacity ratio should be formulated for polyatomic compounds by replacing the monoatomic density of states with a polyatomic formulation in future studies.
Further simulations of gas dynamical processes, both atomistic and continuous, would be necessary to explore the limits of the present theory.
The non-Boltzmann energy distributions in between the Boltzmann and the delta energy distributions are of particular interest, but molecular dynamics simulations will require rather sophisticated thermostating routines for maintaining such non-Boltzmann energy distributions.
Ultimately, experiments need to be designed which would allow to test the presented theoretical and computational findings.

% final conclusion
The present work provides novel insights into gas dynamics of non-Boltzmann gases and proposes interesting and partly counter-intuitive aspects of the heat capacity ratio.
Since the heat capacity ratio is a key property for describing gas dynamical processes, the present findings might help to improve the understanding of gas dynamical phenomena and could potentially allow discovering novel phenomena.

\backmatter

\bmhead{Supplementary information}
Animated sequences of exemplary Boltzmann and non-Boltzmann molecular dynamics simulations, corresponding to the frames presented in Figure~\ref{fig:Sequences}.

%\bmhead{Acknowledgments}

%\bmhead{Data availability} The datasets generated during and/or analyzed during the current study are available from the corresponding author on reasonable request.

\section*{Declarations}

\bmhead{Conflict of interest} The author has no financial or proprietary interests in any material discussed in this article.


\bibliographystyle{unsrt}
%\bibliography{../../../literature}% common bib file
\bibliography{literature}% common bib file


\end{document}